\documentstyle[11pt,aaspp4]{article}
\input tex.def
\slugcomment{Accepted for publication in {\it The Astrophysical Journal}.}
\lefthead{Ho}
\righthead{Radio Emission in Elliptical and S0 Galaxies}

\begin{document}

\title{What Powers the Compact Radio Emission in Nearby \ 
Elliptical and S0 Galaxies?}

\author{Luis C. Ho\footnote{Current address: Carnegie Observatories,
813 Santa Barbara St., Pasadena, CA 91101-1292.}}
\affil{Harvard-Smithsonian Center for Astrophysics, 60 Garden St., Cambridge, 
MA 02138}

\begin{abstract}

Many nearby early-type (elliptical and S0) galaxies contain
weak (milli-Jansky level) nuclear radio sources on scales a few hundred 
parsecs or less.  The origin of the radio emission, however, has remained 
unclear, especially in volume-limited samples that select intrinsically less 
luminous galaxies.  Both active galactic nuclei and nuclear star formation 
have been suggested as possible mechanisms for producing the radio emission.
This paper utilizes optical spectroscopic information to address this issue.  
A substantial fraction of the early-type galaxies surveyed with the Very Large 
Array by Wrobel \& Heeschen (1991) exhibits detectable optical emission lines 
in their nuclei down to very sensitive limits.  Comparison of the observed 
radio continuum power with that expected from the thermal gas traced by the 
optical emission lines implies that the bulk of the radio emission 
is nonthermal.  

Both the incidence and the strength of optical line emission correlate with 
the radio power.  At a fixed line luminosity, ellipticals have stronger 
radio cores than S0s.  The relation between radio power and line emission 
observed in this sample is consistent with the low-luminosity extension of 
similar relations seen in classical radio galaxies and luminous Seyfert 
nuclei.  A plausible interpretation of this result is that the weak nuclear 
sources in nearby early-type galaxies are the low-luminosity counterparts of 
more powerful AGNs.  The spectroscopic evidence supports this picture.  
Most of the emission-line objects are optically classified as Seyfert nuclei 
or low-ionization nuclear emission-line regions (LINERs), the majority of 
which are likely to be accretion-powered sources.

\end{abstract}


\section{Motivation}
The effort to investigate low-level nuclear activity in nearby 
galaxies has proceeded along two separate, largely independent routes.  
The more direct and traditional approach utilizes optical spectroscopy of 
magnitude-limited samples, usually without highly restrictive selection 
by Hubble type (see Ho 1996 for a review).  These surveys arrive at the 
consensus that mild nuclear activity attributable to the phenomenon of active 
galactic nuclei (AGNs) is very common in nearby galaxies, particularly those 
with a prominent bulge component.   Along similar lines, sensitive radio 
interferometric observations have revealed intrinsically weak nuclear sources 
in an increasingly large number of purportedly ``normal'' galaxies (e.g., 
Heeschen 1970; Ekers \& Ekers 1973; Crane 1979; van der Hulst, Crane, \& Keel 
1981; Condon \etal 1982).  To date, the most comprehensive surveys for weak, 
compact nuclear radio emission have concentrated on elliptical and S0 
galaxies.  (As a working definition, ``compact'' here denotes angular scales 
\lax 5\asec, which, for the relatively nearby galaxies addressed in this 
paper, pertain to physical dimensions of several hundred parsecs.)  The two 
main studies are those by Sadler \etal (1989) and Wrobel \& Heeschen (1991, 
hereafter WH), which selected southern and northern sources, respectively; 
both employed the Very Large Array (VLA) at an observing frequency of 5~GHz 
(6~cm), which resulted in a synthesized beam with an angular resolution of 
~$\sim$5\asec.  Both of these studies achieved sensitivities well below 1~mJy 
and detected a large number of previously unknown, low-power radio cores 
associated with the galaxy nuclei.  Some of the newly discovered sources have 
radio powers as low as 10$^{19}$ W Hz$^{-1}$ at 5~GHz, many orders of magnitude 
less than the cores of classical radio galaxies or quasars.

What is the physical nature of these low-power radio cores in nearby 
early-type galaxies?  Are they truly scaled-down versions of distant, 
more powerful AGNs, or can they be explained by less exotic processes? 
Sadler \etal (1989) favored the AGN interpretation on the basis of optical
spectroscopic information available for their sample.  Subsequent radio 
observations by Slee \etal (1994) of a radio-bright subset at a much higher 
angular resolution revealed that most of the radio emission originates 
from extremely small, parsec-scale cores.  The high brightness temperatures 
inferred ($T_B\,>\,10^5$ K) and the preponderance of flat radio spectra 
support a nonthermal origin for the emission, at least for the brightest 
sources they selected.  The situation for the northern objects, however, is 
less clear, for, until recently, optical spectra were not widely available for 
the WH sample.  And unlike the southern sample, there are yet no 
systematic follow-up radio observations with sufficient angular resolution to 
yield meaningful constraints on the brightness temperature, nor are there,
in general, spectral indices to diagnose the emission mechanism.  The 
WH sample contains fewer high-luminosity galaxies than the sample of 
Sadler et al. because the former is selected both by magnitude and by distance.
Since the physical origin of nuclear activity depends on galaxy luminosity 
(e.g., Phillips \etal 1986; Ho, Filippenko, \& Sargent 1997b), the results 
of the southern survey cannot be readily generalized to the northern survey.
Indeed, WH argued, on the basis of an observed correlation 
between the radio and the far-infrared emission, that star formation, not 
AGN activity, predominantly powers the nuclei in their sample.

The purpose of this paper is to clarify the nature of the radio nuclei in 
the WH sample of early-type galaxies with the aid of a newly completed optical
spectroscopic survey of nearby galaxies.  We will argue that the bulk of the 
radio emission arises from accretion-powered activity, as was proposed by 
Sadler \etal (1989) for the southern counterpart of the WH survey.  

\section{Optical Emission-Line Properties}

The WH sample formally comprises 216 elliptical and S0 galaxies ($T\,\leq$ 
--1) in the Center for Astrophysics redshift survey (Huchra \etal 1983) that 
satisfy $B\,<$ 14 mag and $cz\,\leq$ 3000 \kms\ (see Wrobel 1991).  This 
sample naturally overlaps heavily with the Palomar optical spectroscopic 
survey of Ho, Filippenko, \& Sargent (1995, 1997a), which selected galaxies 
from the Revised Shapley-Ames Catalog of Bright Galaxies (Sandage \& Tammann 
1981) having $B_T\,\leq$ 12.5 mag and declinations $>$0\deg.  The Palomar 
survey contains 145 E and S0 galaxies, and 109 of these (38 E and 71 S0) are in 
common with the WH sample.  Figure 1 shows the fractional radio luminosity 
function of the 109 objects calculated using the Kaplan-Meier estimator 
(Feigelson \& Nelson 1985) to take into account the significant number of 
upper limits (69) present.  The fractional luminosity 
functions of ellipticals and S0s are also shown separately to reinforce 
the well-known fact (e.g., Sadler \etal 1989) that the strength of the 
radio source increases with the size of the bulge component.  The 
mean 5-GHz radio power is (2.1$\pm$1.3)\e{21} W Hz$^{-1}$ for the entire 
sample, (5.5$\pm$3.7)\e{21} W Hz$^{-1}$ for the ellipticals alone, and
(0.3$\pm$0.1)\e{21} W Hz$^{-1}$ for the S0s alone.

Ho \etal (1997a) give a variety of spectroscopic parameters measured for the 
nuclear regions; the angular scale of these measurements 
(2\asec$\times$4\asec) roughly matches that of the radio synthesized beam. For 
the present purpose, the most pertinent of the parameters are the spectral 
classifications of the nuclei and their luminosities as measured in the H\al\ 
emission line\footnote{Ho \etal (1997a) tabulated emission-line fluxes and 
luminosities for only about 80\% of the survey galaxies because the rest were 
observed under nonphotometric conditions.  The present analysis utilizes 
additional line fluxes gathered from the literature; the supplementary data 
will be presented in a forthcoming paper.}.  The following analysis uses the 
distances given in Ho \etal (1997a), which are based on $H_{\rm 0}$ = 75 \kms\ 
Mpc$^{-1}$.  As described by Ho \etal (1997a, 1997b), the Palomar survey 
distinguishes four classes of emission-line nuclei: \hii\ nuclei, Seyfert 
nuclei, low-ionization nuclear emission-line regions (LINERs; Heckman 1980), 
and transition objects (composite LINER+\hii\ nucleus).  A minority of nuclei 
do not have detectable emission lines down to an equivalent-width limit of 
$\sim$0.25 \AA.  The primary spectral classification system is based solely on 
the relative intensities of the narrow optical emission lines.  \hii\ nuclei, 
whose spectra closely resemble those of \hii\ regions and are therefore 
assumed to be photoionized by young, massive stars, have relatively weak lines 
of \oi\ \lamb\lamb6300, 6363, \nii\ \lamb\lamb6548, 6583, and \sii\ \lamb\lamb 
6716, 6731 (compared to, say, H\al).   The other three groups, which represent 
variants of AGNs, are recognized by their exceptionally strong low-ionization 
lines of \oi, \nii, and \sii.  LINERs differ from Seyferts by their level 
of excitation, as measured by the ratio of \oiii\ \lamb5007 to H\bet, and 
transition objects have spectra intermediate between those of LINERs and 
\hii\ nuclei.  Some members within each of the AGN classes show evidence 
for a weak broad (widths of several thousand \kms) H\al\ emission line (Ho 
\etal 1997c), and these are designated ``type 1'' objects.

Table 1 gives a breakdown of the spectral classification of the objects 
in common with the WH survey.  The first point to notice is that a substantial
fraction of the galaxies (61\%) have detectable emission lines, in agreement 
with the high detection rate reported by Phillips \etal (1986) for the 
southern sample.  The somewhat higher detection rate of the Palomar sample 
results from its greater sensitivity and the smaller average distance of the 
objects.  Second, nuclei with emission lines are at least ten times more 
likely to be detected in the radio than those without emission lines.  The 
fraction of emission-line nuclei detected in the radio is 57\%, compared 
with 5\% for nuclei not detected in emission, a difference significant at a 
level of $>$99.99\% according to the $\chi^2$ for the 2$\times$2 contingency 
table for these results.
%
%
Expressed in another way, of the 40 nuclei detected in the radio, 
38 (95\%) have optical emission lines, whereas among the 69 nuclei with 
radio upper limits, only 29 (42\%) do.  
%
%
Galaxies detected as radio sources, therefore, have a much higher likelihood 
of being detected in optical emission lines, and vice versa.  This is further 
reflected in the large difference between the strengths of the emission lines: 
the galaxies detected as radio sources have an average extinction-corrected 
H\al\ luminosity (properly accounting for upper limits), of 
(4.2$\pm$1.1)\e{39} \lum, one order of magnitude higher than the 
radio-undetected galaxies, which have an average $L$(H\al) = 
(3.9$\pm$0.9)\e{38} \lum.


Most revealing, however, is the fact that the vast majority of the 
emission-line nuclei (59/67 or $\sim$90\%) are spectroscopically classified as 
AGNs.  
The active nuclei are extremely faint in comparison to traditionally-studied 
AGNs (Fig. 2); the average $L$(H\al) is only 
$\sim$2\e{39} \lum.  The elliptical galaxies exhibit lower levels of optical 
line emission than the S0s [average $L$(H\al) = 0.8\e{39} \lum\ vs. 2.5\e{39} 
\lum], probably reflecting a difference in the central gas content between the 
two morphological types.  The typical electron density of the emission-line 
regions in these objects, as estimated from the \sii\ \lamb\lamb 6716, 6731 
doublet, is 200--300 \cc, which, when combined with the H\al\ luminosities, 
imply an ionized hydrogen mass of $\sim$10$^4$ \solmass.  Most of the AGNs 
belong to the low-ionization category, namely LINERs and LINER/\hii\ 
composites, as is the case in spiral galaxies (Ho \etal 1997b).  The fraction 
of AGNs occupied by LINERs among elliptical and S0 galaxies, 85\%, is somewhat 
higher than among spirals, which is $\sim$70\% (Ho \etal 1997b).  The 
preponderance of LINERs in galaxies with the earliest Hubble types and the 
tendency for the most compact radio cores to be found in bulge-dominated 
systems (e.g., Sadler \etal 1995) no doubt leads to the common suggestion, 
first made by Heckman (1980), that there exists a statistical connection 
between LINERs and compact radio sources.  Indeed, Disney \& Cromwell (1971) 
had remarked, before Heckman coined the term ``LINER,'' on the unusually low 
ionization state of the spectra in their sample of elliptical galaxies with 
radio-bright nuclei.  However, LINERs certainly exist in more diverse 
environments than just early-type galaxies, and it should be stressed that 
a preferred association between compact radio cores and the LINER phenomenon 
has not yet been established unambiguously.  Hummel \etal (1990), who 
analyzed a galaxy sample with a more representative mixture of Hubble types, 
did {\it not} find a statistically higher incidence of compact radio emission 
in LINERs compared to \hii\ nuclei.  In the present sample, the radio 
detection rate in \hii\ nuclei (50\%) is also very similar to that in AGNs 
(58\%); furthermore, the radio detection rate does not appear to depend on 
the AGN class.  The statistics based on our sample alone should be taken as 
tentative because of the small number of \hii\ nuclei (8) and Seyferts (9) 
available.  When combined with the findings of Hummel et al., however, they 
strongly suggest that the presence or absence of nuclear radio emission, at 
least on a scale of several hundred parsecs, is {\it not} a reliable 
discriminator between AGNs and star-forming nuclei or between various optical 
classes of AGNs.  Observations at higher angular resolution may be needed to 
better isolate a central point source, if present, and spectral information 
may help to identify the emission mechanism.

\hii\ nuclei rarely exist in elliptical galaxies; none, in fact, are found in 
the present sample or in the entire Palomar survey (Ho \etal 1997b).  Likewise,
the nuclei in the vast majority of the S0 galaxies are also optically 
identified with AGNs rather than \hii\ nuclei.  The eight S0 galaxies with 
\hii\ nuclei in Table 1, which comprise less than 20\% of the S0s with 
emission-line nuclei, stand out as having somewhat lower optical luminosities 
(by about 1 mag) than those classified as AGNs.  Thus, insofar as the 
optical spectral classification of a nucleus gives a reliable indication
of its dominant central energy source, our results, in conjunction with 
those of Sadler \etal (1989), indicate that nonstellar activity from AGNs, 
not star formation, powers the weak radio cores commonly found in nearby 
early-type galaxies.


\section{Relationship Between Radio Continuum and Optical Line Emission}

Several older studies have reported an association of compact radio emission 
with the presence of optical emission lines in the centers of early-type 
galaxies (Disney \& Cromwell 1971; Ekers \& Ekers 1973; O'Connell \& Dressel 
1978).  The sample analyzed in this paper supports this result with the 
finding that both the incidence and the strength of optical emission lines 
are correlated with the presence of radio emission.  Although there is a large 
range of radio power ($P_{\rm 5GHz}$) at any fixed H\al\ luminosity 
[$L$(H\al)], the two quantities appear correlated when the upper limits are 
properly taken into account (Fig. 3); the correlation holds for ellipticals 
and S0s individually and for both types combined.  The relation between 
$P_{\rm 5GHz}$ and $L$(H\al) for the ellipticals appears offset and 
possibly steeper than that for the S0s.  For a given line luminosity, 
ellipticals have stronger radio sources than S0s do, although the spread within 
each morphological type and the overlap between the two are considerable.
The error bars in the lower right corner of the diagram give an estimate of 
the typical uncertainty associated with the data, although three additional 
factors most likely further contribute to the scatter in the plot.  First, 
the core radio emission is likely to be at least moderately variable.  Second,
as discussed at length in Ho \etal (1997a), some individual H\al\ 
measurements in the Palomar survey can be quite uncertain because of 
slit losses and imperfect photometric calibration.
And finally, although the H\al\ luminosities have been corrected for 
Galactic and internal extinction, the latter contribution is not always 
unambiguous in weak emission-line objects because the Balmer decrement, 
on which the correction depends, can be difficult to determine accurately.

As emphasized by Sadler \etal (1989), it is nontrivial to assess the 
statistical relation between $P_{\rm 5GHz}$ and $L$(H\al) because 
each quantity is itself correlated with the distance and with the total 
optical luminosity of the galaxy.  A possible correlation between 
$P_{\rm 5GHz}$ and $L$(H\al) in the presence of one of these third variables 
was evaluated using the test for partial correlation with censored data 
described by Akritas \& Siebert (1996).  Choosing the total blue absolute 
magnitude as the test variable, we find that $P_{\rm 5GHz}$ correlates very 
significantly with $L$(H\al). The partial Kendall's $\tau$ 
coefficient is 0.299 and the variance is 0.0432 (see Akritas \& Siebert), 
which imply that the null hypothesis that $P_{\rm 5GHz}$ and $L$(H\al) are 
uncorrelated can be rejected at a significance level of $< 10^{-6}$; the
same holds if distance becomes the test variable.  In both cases the 
correlation is stronger for S0s than for ellipticals, but the correlation 
for the latter is still highly significant.

These results conflict with those of Sadler \etal (1989), who concluded 
that early-type galaxies with radio cores do not have a higher 
incidence of optical emission lines than those without radio cores and 
that radio power is not correlated with optical emission-line 
luminosity.   These authors used a likelihood ratio test based on a 
maximum-likelihood estimate of the bivariate luminosity function of 
$P_{\rm 5GHz}$ and $L$(\nii), where $L$(\nii) is the luminosity of the 
\nii\ \lamb 6583 emission line, and found that the two variables are not 
strongly correlated, even {\it without} accounting for the known partial 
correlation of each with a third variable (distance or total galaxy 
luminosity).  To try to understand the root of this discrepancy
\footnote{Note that in the present sample of galaxies the correlation between
radio power and optical emission-line luminosity remains nearly as significant
if $L$(\nii) is substituted for $L$(H\al).  We chose to use $L$(H\al) instead 
of $L$(\nii) because a luminosity based on a recombination line has a more 
straightforward physical interpretation than a collisionally-excited line.}, 
the data of Sadler et al.  were reanalyzed using the same statistical method 
employed here.  Table 4 of Sadler et al. lists 167 objects having both radio 
and optical data, with 123 upper limits for $P_{\rm 5GHz}$ and 76 upper limits 
for $L$(\nii).  The conventional generalized Kendall's $\tau$ test (Isobe, 
Feigelson, \& Nelson 1986) yields a correlation coefficient of 0.189 and 
a significance level of 4\e{-5}.  Thus, contrary to the conclusion of Sadler 
et al., $P_{\rm 5GHz}$  and $L$(\nii) {\it are} highly correlated.  
The same conclusion holds even after allowing for dependence on a third 
variable. Taking the total absolute magnitude of the galaxy again as the 
test variable, the partial Kendall's $\tau$ test formally yields a rather small 
correlation coefficient of 0.0679 (variance = 0.0223), but it is still 
statistically significant (the null hypothesis can be rejected with a 
probability of 0.2\%).  It is unclear why the statistical method used by 
Sadler et al. gives an inconsistent result.  Akritas (1997) 
notes that the asymptotic distribution that Sadler et al. assume for their 
likelihood ratio statistic (equation B8 in their Appendix B) is known to be 
formally valid only for uncensored data; its behavior in the presence of 
censoring is unknown.

The majority of the radio sources in the WH sample do not yet have published 
radio spectral indices or meaningful constraints on their brightness 
temperatures to enable discrimination between a thermal or nonthermal origin 
for the radio emission.  In this regard, it is instructive to consider a 
constraint that can be derived from the strength of the H\al\ emission, which 
is sampled roughly over the same region as the
VLA measurements.  A purely thermal source with an electron temperature of
10$^4$ K generates approximately 10$^{-13}$ $L$(H\al) W Hz$^{-1}$ of
radio power at 5~GHz (see, e.g., Ulvestad, Wilson, \& Sramek 1981),
with $L$(H\al) expressed in units of W.  Figure 4 illustrates that the vast
majority of the sources fall well above this threshold, which implies that
thermal emission contributes negligibly to the radio continuum at this
frequency.  This is consistent with the hypothesis that the nuclei are AGNs,
although such an interpretation is not required because supernova remnants can
generate nonthermal radio emission.  The objects that lie near the
threshold are all S0 galaxies, most being systems of fairly low luminosity
($M^0_{B_T}$ \gax --19.5 mag).  Several of the S0s with \hii\ nuclei, notably,
have radio powers consistent with the thermal limit, although not all the
objects near this limit are classified as \hii\ nuclei.  More generally,
there exists a loose correlation between the radio ``excess'' and the
galaxy luminosity.  The objects with the largest ratio of radio to
optical line luminosity tend to be the most luminous, massive galaxies,
although there is a considerable spread of this ratio at any given total
luminosity. (As discussed above in connection with Figure 3, the intrinsic
scatter is likely to be smaller than indicated because of observational
factors.)

\section{Relation of Nearby Early-Type Galaxies to More Powerful AGNs}

The positive correlation between radio power and optical emission-line 
luminosity in itself, however, does not point to an obvious physical 
explanation for such a relationship.  A correlation between 20-cm radio 
power and the luminosity of the \oiii\ \lamb 5007 forbidden line has been 
recognized for quite some time in Seyfert galaxies (de Bruyn \& Wilson 1978; 
Whittle 1992b), most of which have much higher luminosities than the objects 
considered here, but even in those objects the physical connection between 
the two variables is not obvious.  Whittle (1992b) finds that both radio power 
and emission-line luminosity are strongly coupled to the bulge luminosity, 
but a residual correlation remains, as is the case here.  Powerful radio 
galaxies also exhibit a correlation between radio power and optical 
emission-line luminosity (Hine \& Longair 1979; Rawlings \etal 1989; Baum \& 
Heckman 1989b; Baum, Zirbel, \& O'Dea 1995; Zirbel \& Baum 1995).  It is 
interesting to consider whether the low-luminosity nuclei in our sample 
extend the radio-power/line-luminosity correlations previously established 
for the more powerful AGNs.  Specifically, one might expect the 
radio-quiet ellipticals in our sample to track the radio galaxies because 
they latter are predominantly giant ellipticals (e.g., Ledlow \& Owen 1995; 
Zirbel 1996); the SOs, on the other hand, might follow the Seyfert population, 
whose host galaxies tend to be bulge-dominated disk systems (e.g., 
Whittle 1992a; Ho \etal 1997b).

The best-fitting relations between radio power and line luminosity found by 
Zirbel \& Baum (1995) are plotted on Figure 3.  Radio galaxies of Fanaroff 
\& Riley (1974) ``class I'' and ``class II''
obey slightly different relations, and both are shown.  The line luminosities 
used by Zirbel \& Baum (1995) represent the sum of the H\al\ and the 
\nii\ \lamb\lamb 6548, 6583 lines.  To translate their results for this 
comparison, we assume that \nii\ \lamb\lamb 6548, 6583 = 2.3H\al, the median 
value observed in elliptical and S0 galaxies by Ho \etal (1997a).  The 
best-fitting lines shown are those for the core radio power, since in 
the WH sample most of the radio emission comes from a compact core, and they 
were derived from objects with $z\,<\,0.5$.  Note that, in the units 
used in Figure 3, the objects in the Zirbel \& Baum sample have 
log $P_{\rm 5 GHz}\,\approx$ 21--27 and log $L$(H\al) $\approx$ 38.5--44.5.
We have also include in the plot the sample of Seyfert nuclei compiled by 
Whittle (1992a); most of these objects are significantly more luminous 
than those in our sample.  Again, for comparison with our data, we computed 
$P_{\rm 5 GHz}$ from Whittle's 1.4-GHz flux densities assuming 
$S_{\nu}\,\propto\,\nu^{-0.7}$, and we calculated $L$(H\al) from his 
\oiii\ luminosities by assuming \oiii\ \lamb5007/H\bet\  = 10 and 
H\al/H\bet\ = 3.1, values typical of Seyfert nuclei (e.g., Shuder \& 
Osterbrock 1981).

Making allowance for the large scatter in the diagram, the nearby radio-quiet 
ellipticals broadly follow the faint end of the relations established by the 
radio galaxies.  The SOs, on the other hand, appear to join fairly well 
with the faint end of the Seyfert galaxy sequence, which, as was noticed by 
Baum \& Heckman (1989b), is distinctly offset from the radio galaxy sequence.
A simple interpretation of these results is that the quiescent nuclei in 
nearby early-type galaxies are the low-luminosity counterparts of the 
more distant, more powerful AGNs.  This interpretation is consistent 
with the spectroscopic evidence presented in \S\ 2.  The spectra of the 
nuclei, with only a few exceptions, decidedly do not resemble the spectra of 
\hii\ regions; instead, most of them look like those of LINERs.  
If one adopts the viewpoint that most LINERs represent another, and perhaps 
the most common, manifestation of the AGN phenomenon (e.g., Halpern \& Steiner 
1983; Ferland \& Netzer 1983; Ho, Filippenko \& Sargent 1993; see reviews 
by Filippenko 1996 and Ho 1998), one would conclude that the weak, compact 
radio emission in the centers of nearby early-type galaxies derives from 
physical processes similar to those operating in more powerful AGNs.  In turn, 
the finding that many nearby early-type galaxies do, in fact, contain 
high-brightness temperature, flat-spectrum radio cores (Slee \etal 1994) 
provides strong support for the AGN-like nature of LINERs, at least for those 
found in elliptical and S0 galaxies.

\section{Conclusions}

We have used optical spectroscopic information to interpret the nature of the 
compact radio emission in a sample of nearby early-type (elliptical and S0) 
galaxies surveyed with the VLA by Wrobel \& Heeschen (1991).  Many of these 
galaxies have weak radio sources (1--few mJy at 5~GHz) on scales of several 
hundred parsecs or less, some with radio powers as low as 10$^{19}$ W Hz$^{-1}$.
The radio sources in ellipticals are more luminous than those in S0s because 
the radio power increases with the optical luminosity of the bulge component.

A substantial fraction of the galaxies ($\sim$60\%) show detectable levels of 
optical nebulosity down to very sensitive limits ($\sim$0.25 \AA\ equivalent 
width for the H\al\ emission line) over the same physical scale probed by the 
radio observations.  The measured H\al\ luminosities, 10$^{38}$--10$^{40}$ 
\lum, imply the presence of 10$^3$--10$^5$ \solmass\ of warm (10$^4$ K) 
ionized hydrogen.  Among S0 galaxies, the quantity of ionized gas 
increases with galaxy luminosity (or size), but, as a class, ellipticals have 
less ionized gas than S0s.   The amount of thermal gas implied by the 
H\al\ measurements falls short of producing, by large factors in most 
cases, the observed radio continuum luminosity.  Most of the radio emission 
is therefore nonthermal.  

Both the incidence and the strength of optical line emission correlate with 
the radio power.  At a fixed line luminosity, ellipticals have stronger 
radio cores than S0s.  The relation between radio power and line emission 
in the nearby radio-quiet ellipticals appears to be an extension of a 
similar relation seen in powerful radio galaxies.  The S0s, on the other 
hand, follow the faint end the correlation established by luminous Seyfert 
galaxies.  It is suggested that the weak nuclear sources seen in nearby 
early-type galaxies are simply the low-luminosity counterparts of more 
distant, luminous AGNs.  The spectroscopic evidence supports this 
interpretation.  The vast majority of the objects with detectable emission 
lines emit optical spectra that differ substantially from those of \hii\ 
regions.  Instead, most of them are classified as LINERs, a few as 
Seyferts.  If LINER nuclei are predominantly accretion-powered sources, as 
suggested by some studies, then most of the radio nuclei can be identified 
as AGNs.

\acknowledgments
The author is supported by a postdoctoral fellowship from the
Harvard-Smithsonian Center for Astrophysics and by NASA grants GO-06837.01-95A
and AR-07527.02-96A from the Space Telescope Science Institute (operated by
AURA, Inc., under NASA contract NAS5-26555).  I thank M. Akritas and the 
Statistical Consulting Center for Astronomy at Penn State University for 
guidance on some statistical issues.  The referee, John Stocke, offered 
constructive criticisms and helpful suggestions.  This work has made use of 
the NASA/IPAC Extragalactic Database (NED) which is operated by the Jet 
Propulsion Laboratory, California Institute of Technology, under contract 
with the National Aeronautics and Space Administration.  

\bigskip


\centerline{\bf{References}}
\medskip

\refindent 
Akritas, M.~G. 1997, private communication

\refindent 
Akritas, M.~G., \& Siebert, J. 1996, \mnras, 278, 919

\refindent 
Baum, S.~A., \& Heckman, T.~M. 1989a, \apj, 336, 681

\refindent 
Baum, S.~A., \& Heckman, T.~M. 1989b, \apj, 336, 702

\refindent 
Baum, S.~A., Zirbel, E.~L., \& O'Dea, C.~P. 1995, \apj, 451, 88

\refindent 
Condon, J.~J., Condon, M.~A., Gisler, G., \& Puschell, J.~J. 1982, \apj, 252,
102

\refindent 
Crane, P.~C. 1979, \aj, 84, 281

\refindent 
de Bruyn, A.~G., \& Wilson, A.~S. 1978, \aa, 64, 433

\refindent 
Disney, M.~J., \& Cromwell, R.~H. 1971, \apj, 164, L35

\refindent 
Ekers, R.~D., \& Ekers, J.~A. 1973, \aa, 24, 247

\refindent 
Fanaroff, B.~L., \& Riley, J.~M. 1974, \mnras, 167, 31P

\refindent 
Feigelson, E.~D., \& Nelson, P.~I. 1985, \apj, 293, 192

\refindent 
Ferland, G. J., \& Netzer, H. 1983, \apj, 264, 105

\refindent 
Filippenko, A.~V. 1996, in The Physics of LINERs in View
of Recent Observations, ed.  M. Eracleous et al.  (San Francisco: ASP), 17

\refindent 
Halpern, J.~P., \& Steiner, J.~E. 1983, \apj, 269, L37

\refindent 
Heckman, T.~M. 1980, \aa, 87, 152

\refindent
Heeschen, D.~S. 1970, Ap. Let., 6, 49

\refindent
Hine, R.~G., \& Longair, M.~S. 1979, \mnras, 188, 111

\refindent
Ho, L.~C. 1996, in The Physics of LINERs in View of
Recent Observations, ed. M. Eracleous et al. (San Francisco: ASP), 103
 
\refindent
Ho, L.~C. 1998, in  The AGN-Galaxy Connection, ed. H.~R. Schmitt, L.~C. Ho, \&
A.~L. Kinney (Advances in Space Research), in press (astro-ph/9807273)

\refindent 
Ho, L.~C., Filippenko, A.~V., \& Sargent, W.~L.~W. 1993, \apj, 417, 63 

\refindent 
Ho, L.~C., Filippenko, A.~V., \& Sargent, W.~L.~W. 1995, \apjs, 98, 477

\refindent
Ho, L.~C., Filippenko, A.~V., \& Sargent, W.~L.~W. 1997a, \apjs, 112, 315

\refindent  
Ho, L.~C., Filippenko, A.~V., \& Sargent, W.~L.~W. 1997b, \apj, 487, 568
 
\refindent  
Ho, L.~C., Filippenko, A.~V., Sargent, W.~L.~W., \& Peng, C.~Y. 1997c, \apjs,
112, 391

\refindent  
Huchra, J.~P., Davis, M., Latham, D., \& Tonry, J. 1983, \apjs, 52, 89

\refindent  
Hummel, E., van der Hulst, J.~M., Kennicutt, R.~C., Jr., Keel, W.~C. 1990,
\aa, 236, 333

\refindent  
Isobe, T., Feigelson, E.~D., \& Nelson, P.~I. 1986, \apj, 306, 490

\refindent  
Ledlow, M.~J., \& Owen, F.~N. 1995, \aj, 110, 1959

\refindent  
O'Connell, R.~W., \& Dressel, L.~L. 1978, \nat, 276, 374

\refindent  
Phillips, M.~M., Jenkins, C.~R., Dopita, M.~A., Sadler, E.~M., \&
Binette, L. 1986, \aj, 91, 1062

\refindent  
Rawlings, S., Saunders, R., Eales, S.~A., \& Mackay, C.~D. 1989, \mnras,
240, 701

\refindent  
Sadler, E.~M., Jenkins, C.~R., \& Kotanyi, C.~G. 1989, \mnras, 240, 591

\refindent  
Sadler, E.~M., Slee, O.~B., Reynolds, J.~E., \& Roy, A.~L. 1995, \mnras, 276,
1373

\refindent  
Sandage, A.~R., \& Tammann, G.~A. 1981, A Revised Shapley-Ames Catalog of
Bright Galaxies (Washington, DC: Carnegie Inst. of Washington)

\refindent  
Shuder, J.~M., \& Osterbrock, D.~E. 1981, \apj, 250, 55

\refindent  
Slee, O.~B., Sadler, E.~M., Reynolds, J.~E., \& Ekers, R.~D. 1994, \mnras, 269,
928

\refindent  
Ulvestad, J.~S., Wilson, A.~S., \& Sramek, R.~A. 1981, \apj, 247, 419

\refindent  
van der Hulst, J.~M., Crane P.~C., \&  Keel, W.~C. 1981, \aj, 86, 1175

\refindent  
Whittle, M. 1992a, \apjs, 79, 49

\refindent  
Whittle, M. 1992b, \apj, 387, 121

\refindent  
Wrobel, J.~M. 1991, \aj, 101, 127

\refindent  
Wrobel, J.~M., \& Heeschen, D.~S. 1991, \aj, 101, 148 (WH)

\refindent  
Zirbel, E.~L. 1996, \apj, 473, 713

\refindent  
Zirbel, E.~L., \& Baum, S.~A. 1995, \apj, 448, 521

\clearpage
\begin{figure}
\plotone{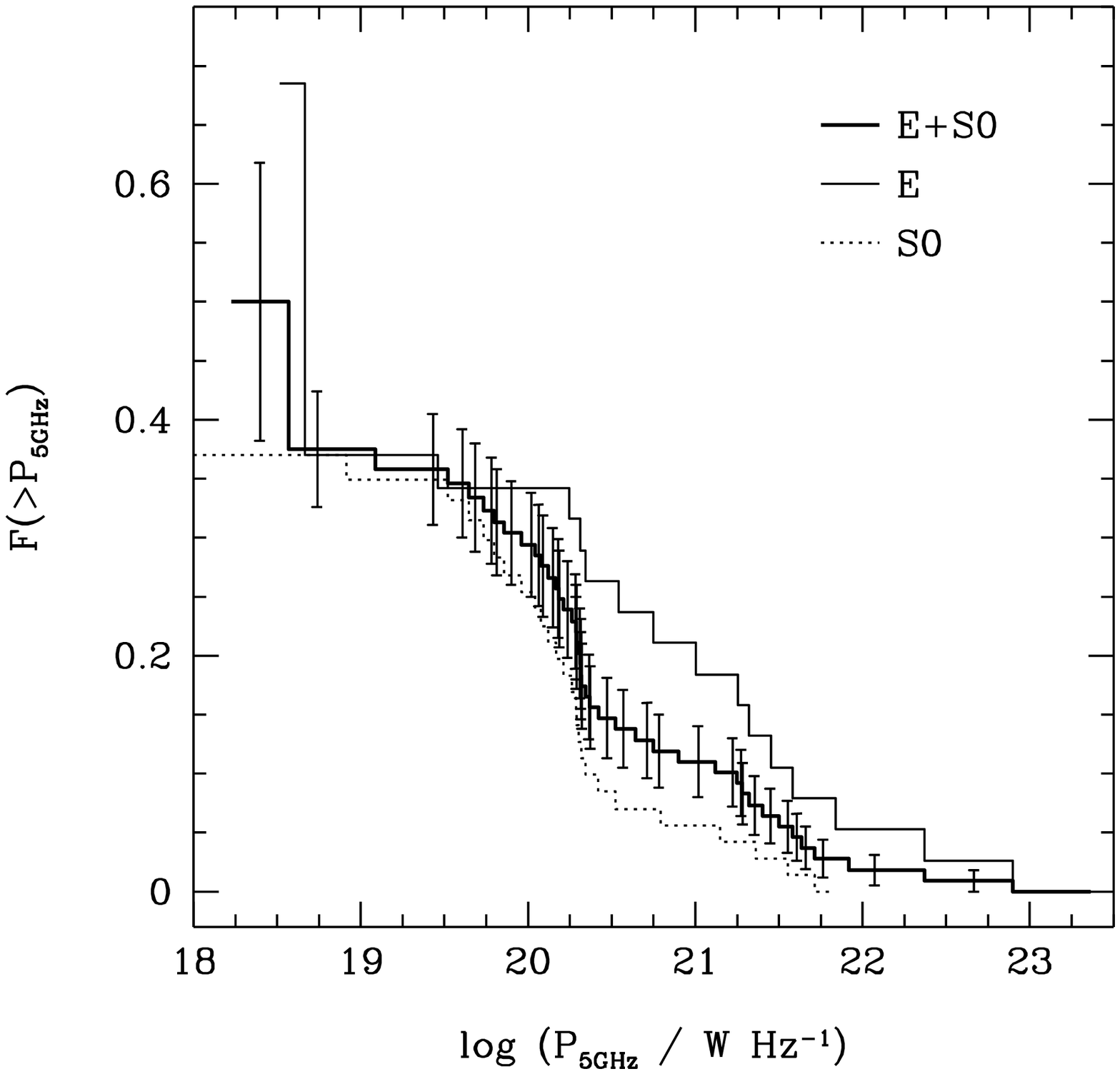}
\caption{
Fractional radio luminosity function for the ellipticals ({\it thin solid
line}), S0s ({\it dotted line}), and ellipticals and S0s combined
({\it thick solid line}).  The error bars pertain to the combined sample.
}
\end{figure}

\clearpage
\begin{figure}
\plotone{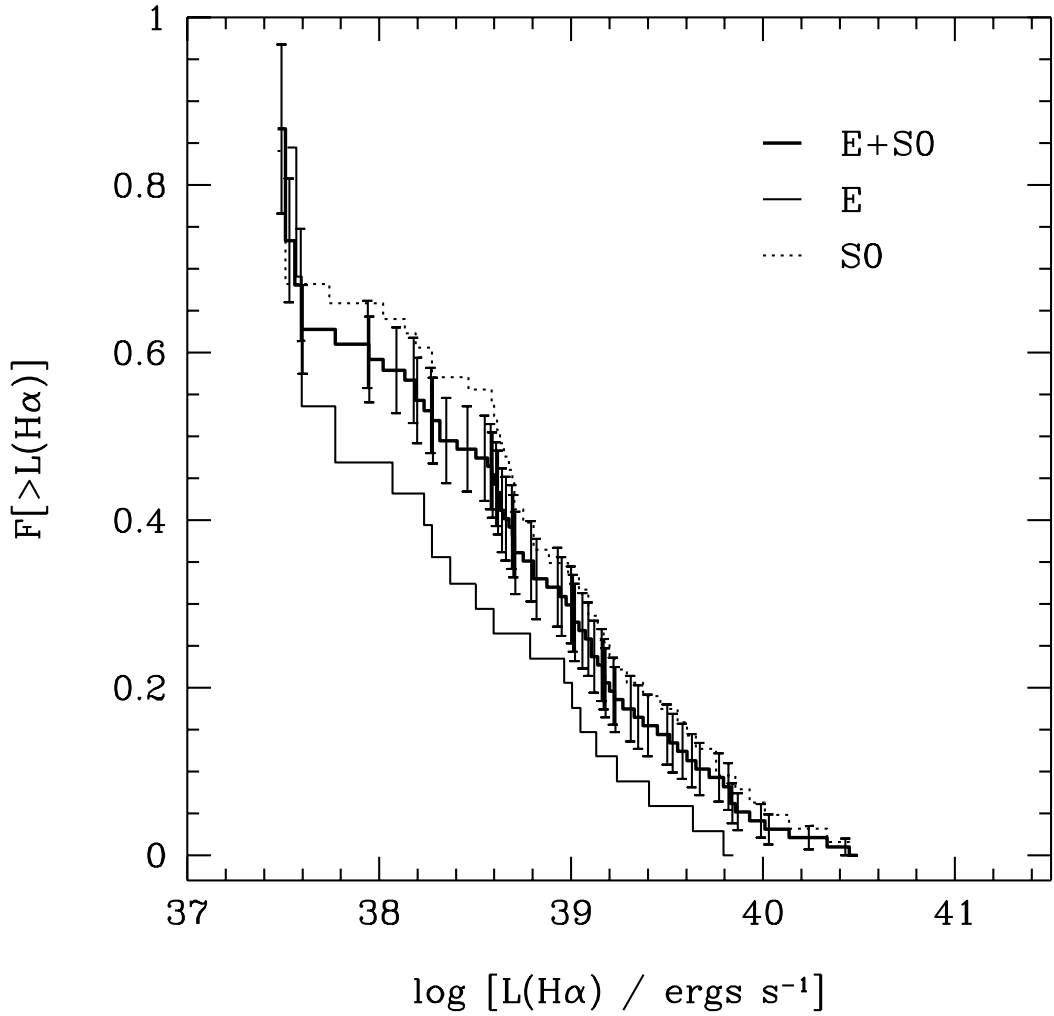}
\caption{
Fractional H\al\ luminosity function for the ellipticals ({\it thin solid
line}), S0s ({\it dotted line}), and ellipticals and S0s combined
({\it thick solid line}).  The error bars pertain to the combined sample,
and the H\al\ luminosities have been corrected for extinction.
}
\end{figure}

\clearpage
\begin{figure}
\plotone{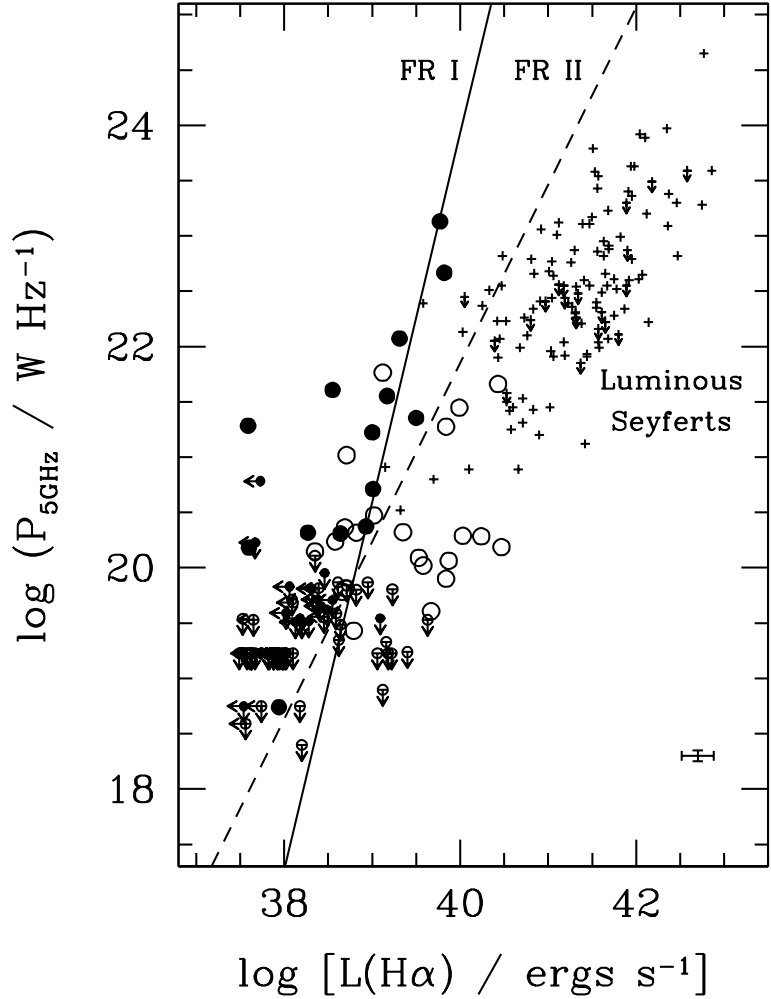}
\caption{
Radio power at 5~GHz versus extinction-corrected H\al\ luminosity.
Elliptical galaxies are shown as {\it filled} symbol, and S0s are plotted
as {\it open} symbols.  The typical uncertainty associated with the data is
illustrated by the error bars in the lower right corner of the diagram.
The sample of luminous Seyfert galaxies compiled by Whittle (1992a) is
shown as {\it plus} symbols (see text for details).  The
best-fitting linear correlations between radio power and emission-line
luminosity for FR~I ({\it solid line}) and FR~II ({\it dashed line}) radio
galaxies were taken from Zirbel \& Baum (1995; see text for details).
Arrows denote upper limits in one or both of the variables.
}
\end{figure}

\clearpage
\begin{figure}
\plotone{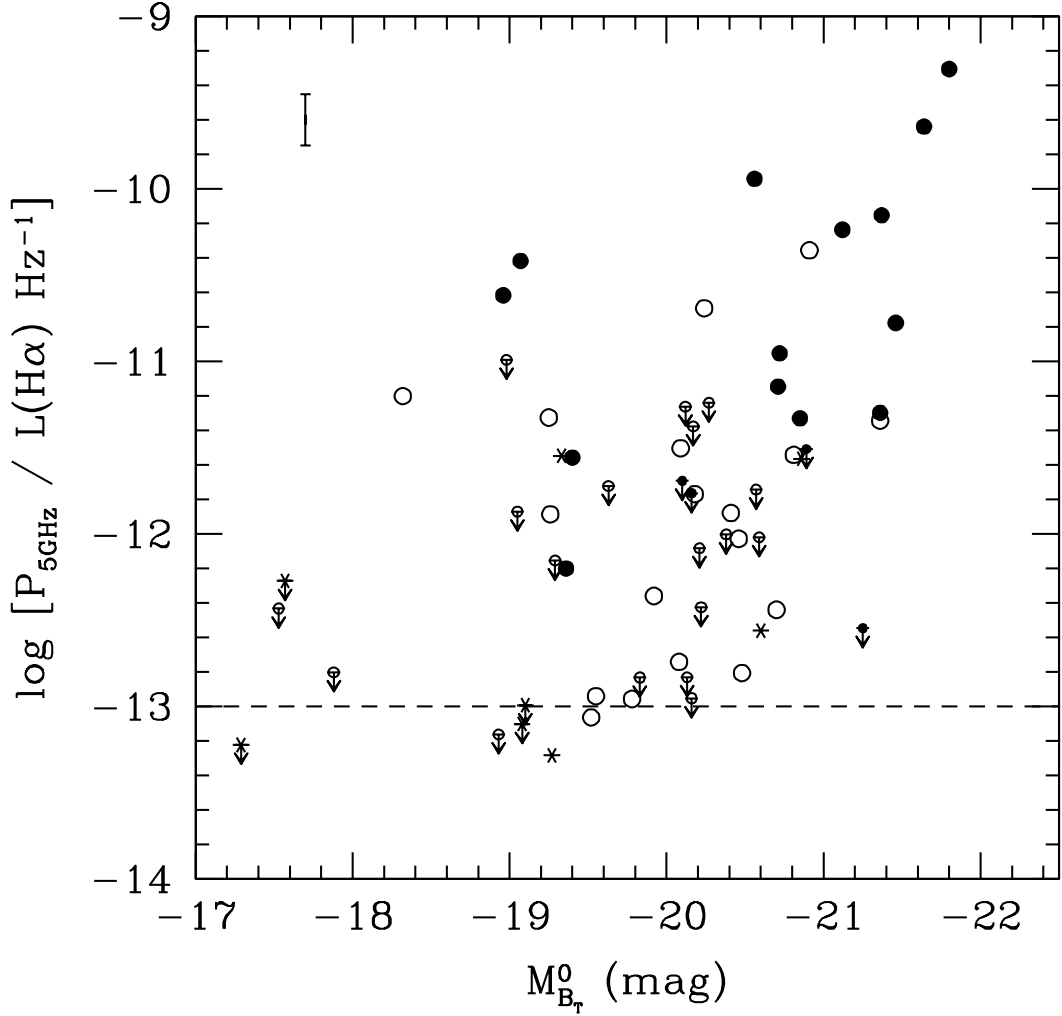}
\caption{The ratio of 5-GHz power to extinction-corrected H\al\ luminosity 
versus the total blue absolute magnitude.  Elliptical galaxies
are shown as {\it filled} symbol, S0s classified as AGNs are plotted as 
{\it open} symbols, and S0s classified as \hii\ nuclei are plotted as 
{\it stars}.  Arrows denote upper limits, and the dashed line indicates
the ratio expected for purely thermal emission.
The typical uncertainty associated with the data is illustrated 
by the error bars in the upper left corner of the diagram.
}
\end{figure}

\clearpage
\begin{figure}
\plotone{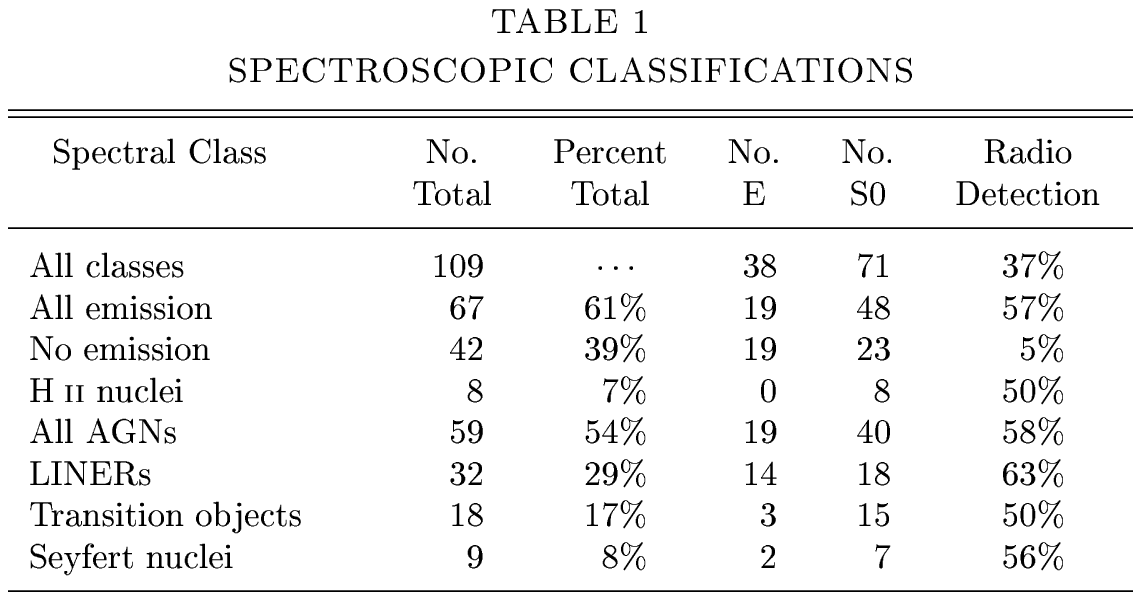}
\end{figure}

\end{document}